\documentclass[10pt, conference,letterpaper]{IEEEtran}

\usepackage{epsfig,amsmath,graphicx,subfigure}
\usepackage{times,rawfonts}
\usepackage{pdfsync}
\usepackage{hyperref}
\IEEEoverridecommandlockouts

\title{LDPC Code Design for the BPSK-constrained Gaussian
  Wiretap Channel}

\author{Chan Wong Wong, Tan F. Wong, and John M. Shea \thanks{The
    authors are with the Wireless Information Networking Group,
    University of Florida, Gainesville, Florida 32611-6130, USA.}}
\begin{document}
\maketitle 

\begin{abstract}
  A coding scheme based on irregular low-density parity-check (LDPC)
  codes is proposed to send secret messages from a source over the
  Gaussian wiretap channel to a destination in the presence of a
  wiretapper, with the restriction that the source can send only
  binary phase-shift keyed (BPSK) symbols.  The secrecy performance
  of the proposed coding scheme is measured by the secret message rate
  through the wiretap channel as well as the equivocation rate about
  the message at the wiretapper. A code search procedure is suggested
  to obtain irregular LDPC codes that achieve good secrecy performance
  in such context.
\end{abstract}

\section{Introduction}\label{sec:intro}
Physical-layer security provides secure communication between desired
users by taking advantages of various physical channel
characteristics. The study of physical-layer security dates back to
Wyner's seminal paper~\cite{Wyner1975}, in which he introduced the
{\it wiretap channel} model. In a wiretap channel, a source tries to
send secret information to a destination in the presence of a
wiretapper.  When the wiretapper channel is a degraded
version of the destination channel, Wyner~\cite{Wyner1975}
described a code design based on group codes such that the source can
transmit messages at a positive rate to the destination.
This degradedness condition was removed in~\cite{Csiszar1978}, which
showed that a positive secrecy rate is possible for the case where the
destination channel is ``less noisy'' than the
wiretapper channel. Generalization of Wyner's work to the
Gaussian wiretap channel was considered
in~\cite{Leung1978}. In~\cite{Ozarow1984}, a code design based on
coset codes was suggested for the type-II (the destination channel is
error free) binary erasure wiretap channel.
Reference~\cite{LiuRuoheng07} considered the design of secure nested
codes for type-II wiretap channels. Recently,
references~\cite{Mahdavifar2010} and \cite{Koyluoglu2010} concurrently
established the result that polar codes~\cite{Arikan2009} can achieve
the secrecy capacity of the degraded binary-input symmetric-output
(BISO) wiretap channels.

Low-density parity-check (LDPC) codes have emerged
as potential practical candidates for providing reliable and secure
communication over wiretap channels. First, LDPC codes provide
excellent ``close-to-capacity'' performance with reasonable
encoding/decoding complexity~\cite{Gal63,MaN96,RichardsonIT01_3}.  
Second, there exists powerful tools
like density evolution~\cite{RichardsonIT01_1} for asymptotic analysis
and code design of LDPC codes.  For example, the authors
of~\cite{Thangaraj2007} constructed LDPC based wiretap codes for
the binary erasure channel (BEC) and the binary symmetric channel (BSC).
In~\cite{Bloch2008}, multilevel coding/multistage decoding using LDPC
codes has been proposed for the quasi-static Rayleigh fading wiretap
channel.  In~\cite{Klinc2009}, punctured LDPC codes were employed in a
coding scheme that aims at reducing the security gap of the Gaussian 
wiretap channel.
In~\cite{Baldi2010}, further reductions in the security gap are
achieved using non-systematic LDPC code obtained by scrambling the
information bits prior to LDPC encoding.

In this paper, we design LDPC codes for sending secret messages over
the Gaussian wiretap channel with binary phase-shift keyed (BPSK)
source symbols. This work is inspired by the results
in~\cite{CWWongTransInfoForensics10} on secret key agreement over the
same wiretap channel, with the availability of an additional public
feedback channel.  In particular, Theorem~2
of~\cite{CWWongTransInfoForensics10} can be modified to show the
existence of regular LDPC code ensembles with increasing block lengths
that achieve secrecy capacity~\cite{Wyner1975,Leung1978} of the
BPSK-constrained Gaussian wiretap channel.  Based on this observation,
we propose a coding scheme which employs irregular LDPC
codes~\cite{RichardsonIT01_2} with finite block lengths to support
practical secret transmission over the Gaussian wiretap channel. The
proposed coding structure allows efficient design of irregular LDPC
codes that give good secrecy performance as measured in terms of
equivocation about the secret message at the wiretapper. This design
constitutes the main technical contribution of the present paper. We
note that codes suggested in~\cite{Klinc2009} can be considered as
unoptimized special cases of the proposed coding scheme. A more
detailed comparison will be given in the sequel.

The outline of the paper is as follows.~\autoref{sec:bpsk_gwc}
introduces the BPSK-constrained Gaussian wiretap channel. The proposed
coding scheme will be discussed in detail
in~\autoref{sec:prop_cod}. In~\autoref{sec:irregldpc}, we describe a
code search algorithm based on density evolution analysis to obtain
good irregular codes for use in the proposed coding scheme. Finally,
conclusions are drawn in~\autoref{sec:con}.

\section{BPSK-constrained Gaussian wiretap
  channel}\label{sec:bpsk_gwc}

We consider the wiretap channel model in which the source tries to
send a secret message to the destination via an additive white
Gaussian noise (AWGN) channel in the presence of a wiretapper. The
wiretapper intends to reconstruct the message by listening to the
transmission through another independent AWGN channel.  The secret
message $M \in \{1,2 \cdots,2^k\}$ is encoded into a transmitted
sequence $X^n = [ X_1, X_2, \cdots , X_n]$. Let $Y^n$ and $Z^n$ denote
the corresponding received sequences at the destination and
wiretapper, respectively.  We restrict the source to transmit only
BPSK symbols, i.e. $X_i \in \{\pm 1\}$\footnote{In later sections,
  whenever appropriate, we implicitly employ the mapping $+1
  \rightarrow 0$ and $-1 \rightarrow 1$, where $0$ and $1$ are the two
  usual elements in GF(2).}.  Then the BPSK-constrained Gaussian
wiretap channel can be modeled as
\begin{equation*}
\begin{split}
Y_i &= \beta X_i + N_i \\
Z_i &= \alpha \beta X_i + \tilde N_i,
\end{split}
\end{equation*}
where $N_i$ and $\tilde N_i$ are independent and identically
distributed (i.i.d.)  zero-mean Gaussian random variables of variance
$\sigma^2$.  Note that $\beta$ is the gain of the BPSK symbols
transmitted by the source.  We impose the source power constraint
$\frac{1}{n} \sum_{j=1}^{n} |X_j|^2 \leq P$ such that $\beta^2 \leq
P$, where $P$ is the maximum power available to the source. Also,
$\alpha$ is a positive constant that models the gain advantage of the
wiretapper over the destination.  Let the (noise) normalized gain be
$\tilde \beta = \beta / \sigma$, then the received signal-to-noise
ratios (SNRs) at the destination and wiretapper are $\tilde \beta^2$
and $\alpha^2 \tilde \beta^2$, respectively.

Assuming a uniform message distribution, the rate of the secret
message is $R_s = \frac{k}{n}$.  Let $\hat{M}$ denote the estimate of
the message at the destination. The level of knowledge of the
wiretapper possesses about the secret message can be quantified by the
equivocation rate $\frac{1}{n}H(M|Z^n)$. A rate-equivocation pair
$(R_s,R_e)$ is achievable if for all $\epsilon > 0$, there exists a
rate-$R_s$ code sequence such that
\begin{enumerate}
\item $\Pr\{ M \neq \hat{M}\} < \epsilon$, and 
\item $R_e < \frac{1}{n}H(M|Z^n) + \epsilon$
\end{enumerate}  
for sufficiently large $n$. When the equivocation rate at the
wiretapper is as large as the secret message rate, i.e. $R_s = R_e$,
we say that the equivocation-rate pair is achievable with {\it perfect
  secrecy}~\cite{Wyner1975}.  The {\it capacity-equivocation region}
of a wiretap channel contains all achievable rate-equivocation pairs
$(R_s,R_e)$.  When $\alpha \leq 1$, specializing the result
in~\cite{Csiszar1978} shows that the capacity-equivocation region of
the BPSK-constrained Gaussian wiretap channel is given by
\begin{eqnarray}\label{eq:region}
  &&0 \leq R_e \leq C_b \nonumber \\ 
  &&R_e \leq R_s \leq C \left(\sqrt{\frac{P}{\sigma^2}}\right), 
\end{eqnarray} 
where 
\begin{equation}\label{eqn:cb}
C_b = \max_{0 \leq \tilde \beta \leq \sqrt{\frac{P}{\sigma^2}}} \Bigg\{
  C(\tilde\beta)-C(\alpha\tilde\beta) \Bigg\},
\end{equation}
and
\begin{equation*}
C(t) = 
  1- \frac{1}{\sqrt{2\pi}} \int_{-\infty}^{\infty} e^{-\frac{(y-t)^2}{2}}
  \log_2 \left ( e^{-2yt} \right ) dy
\end{equation*}
is the channel capacity of AWGN channel with BPSK input.  The {\it
  secrecy capacity} of the wiretap channel is defined as the maximum
secret message rate such that the condition of perfect secrecy is
satisfied.  For the BPSK-constrained Gaussian wiretap channel, the
secrecy capacity is given by $C_b$ if $\alpha \leq 1$.

\begin{figure}
\centering
\begin{center}
\includegraphics[width=0.5\textwidth]{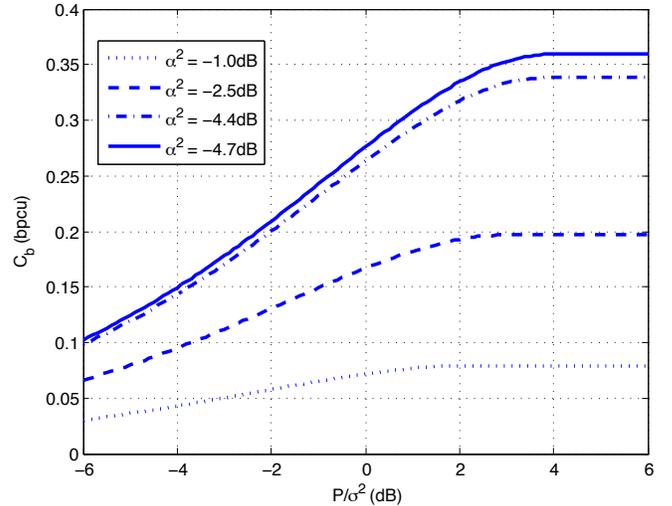}
\end{center}
\caption{The secrecy capacity $C_b$ of the BPSK-constrained Gaussian
  wiretap channel for different value of $\alpha^2$.}
\label{fig:cb}
\vspace*{-10pt}
\end{figure}

We note that $C_b$ is achieved when $X_i$ is equiprobable; but it is
not necessarily achieved by transmitting at the maximum allowable
power $P$.  Fig.~\ref{fig:cb} shows the plot of $C_b$, in units of
bits per (wiretap) channel use (bpcu), versus the
maximum allowable SNR $P/\sigma^2$ for $\alpha^2=-1.0,-2.5,-4.4,$ and
$-4.7$~dB, respectively.

\section{Secret LDPC coding scheme}\label{sec:prop_cod}
In this section, we describe the proposed coding scheme for the
BPSK-constrained Gaussian wiretap channel. Inspired by our previous
work~\cite{CWWongTransInfoForensics10} on secret key agreement over
the same wiretap channel, the proposed coding scheme employs irregular
LDPC codes and its secrecy performance will be evaluated by measuring
the equivocation rate of the secret message at the wiretapper.

Similar to~\cite{CWWongTransInfoForensics10}, we start with an
$(n,l,k)$ \emph{secret binary linear block code} defined by the pair
$(\mathcal{C},\mathcal{W})$, where $\mathcal{C}$ is an $(n,l)$ binary
linear block code with $2^l$ distinct codewords of length $n$ and
$\mathcal{W}$ is an $(l-k)$-dimensional subspace in $\mathcal{C}$.
The ratios $R_c=\frac{l}{n}$ and $R_s=\frac{k}{n}$ will be referred to
as the \emph{code rate} and \emph{secret rate} of
$(\mathcal{C},\mathcal{W})$, respectively.  The pair
$(\mathcal{C},\mathcal{W})$ will be employed to support practical
secret transmission over the wiretap channel.  We take the following
considerations into account when choosing
$(\mathcal{C},\mathcal{W})$. First, the choice of
$(\mathcal{C},\mathcal{W})$ should admit efficient encoder and decoder
in order to make the proposed coding scheme practically implementable.
Second, the choice of $(\mathcal{C},\mathcal{W})$ should avoid
transmitting the secret message $M$ ``directly'' through the channel
since such ``direct'' transmission is undesirable.  Third, the choice
of $(\mathcal{C},\mathcal{W})$ should permit a design procedure based
on which ``good'' codes can be obtained.

To satisfy these considerations, we choose an $(m,l)$ linear block
code $\mathcal{C'}$ from an ensemble of irregular LDPC codes, where
$m=n+k$. Then the pair $(\mathcal{C},\mathcal{W})$ is chosen as
follows. Let $H$ be the parity-check matrix associated with
$\mathcal{C'}$ and assume $H =[A,B]$ where $B$ is an $(m-l)\!\times\!
(m-l)$ lower triangular matrix\footnote{This lower triangular matrix
  can be obtained through performing Gaussian elimination. If
  efficient encoding is necessary, an approximate lower triangular
  matrix as described in~\cite{RichardsonIT01_3} can be used
  instead.}.  Let $\tilde{X}^{m}=[c^k, d^{l-k}, e^{m-l}]$ where
$e^{m-l}=[c^k, d^{l-k}]A^T(B^{-1})^T$ denote a codeword of
$\mathcal{C'}$.
Then the $(n,l)$ linear block code $\mathcal{C}$ is chosen to be set
of codewords obtained by removing $c^k$ from $\tilde{X}^{m}$.  That
is, $\mathcal{C}$ is a punctured version of $\mathcal{C'}$.  The
subspace $\mathcal{W}$ is chosen to be the subset of (punctured)
codewords obtained by setting $c^k$ to zero.  The coding scheme
employing the pair $(\mathcal{C},\mathcal{W})$ is as follows:
\begin{enumerate}
\item{\bf Encoding:} The source sets $c^k$ to be the $k$-bits secret
  message $M$ and chooses $d^{l-k}$ randomly according to a uniform
  distribution.  Then it calculates $e^{m-l}=[c^k,
  d^{l-k}]A^T(B^{-1})^T$ and sends $X^n = [d^{l-k}, e^{m-l}]$ to the
  destination through the Gaussian wiretap channel.
   
\item{\bf Decoding:} The destination performs belief propagation (BP)
  decoding to decode $\tilde{X}^{m}$ using its channel observation
  $Y^n$. The first $k$ bits of the decoded codeword give the estimate
  $\hat{M}$ of the secret message.
\end{enumerate}

We evaluate the secrecy performance of the proposed coding scheme in
the context of achievable rate-equivocation pair defined
in~\autoref{sec:bpsk_gwc}.  First, if the BP decoder at the
destination achieves error probability $\epsilon_d$, then we have
$\Pr\{M \neq \hat M\} \leq \epsilon_d$.
Hence, Condition 1 in~\autoref{sec:bpsk_gwc} is satisfied if
$\epsilon_d$ is small enough.  Second, the uncertainty about the
message $M$ at the wiretapper given his received sequence $Z^n$ is
\begin{eqnarray}\label{eqn:HMZn}
\lefteqn{H(M|Z^n)} \nonumber \\
         &=& H(X^n|Z^n) + H(M|Z^n,X^n) - H(X^n|M,Z^n) \nonumber \\
         &=& H(X^n) - I(X^n;Z^n) + H(M|Z^n,X^n) - \nonumber\\
         && H(X^n|M,Z^n). 
\end{eqnarray}
Based on the memoryless nature of the source-to-wiretapper channel and
the encoding process, we have $I(X^n;Z^n) \leq n
C(\alpha\tilde\beta)$, $H(X^n) = l$~\footnote{This is valid when
  $\mathcal{C}$ contains $2^l$ distinct codewords, which is in turn
  the case with very high probability if $\mathcal{C}'$ is chosen
  randomly in the usual manner described in~\cite{RichardsonIT01_2}.}
and $H(M|Z^n,X^n) \leq H(M|X^n) = 0$, respectively. Moreover, consider
a fictitious receiver at the wiretapper trying to decode for $X^n$
from observing $Z^n$ and $M$. Suppose that the average error
probability achieved by this receiver is $\epsilon_w$. Then we have
$H(X^n|M,Z^n) \leq 1 + (l-k)\epsilon_w$ by Fano's inequality.  Putting
all these back to~\eqref{eqn:HMZn}, we obtain
\begin{equation}\label{eqn:nHMZn}
  \frac{1}{n}H(M|Z^n) \geq
  R_c -  C(\alpha\tilde\beta) - (R_c-R_s) \epsilon_w - \frac{1}{n}. 
\end{equation}
Let $R_e = R_c - C(\alpha\tilde\beta)$, then Condition 2
in~\autoref{sec:bpsk_gwc} is satisfied if $\epsilon_w$ is small enough
and $n$ is large enough. Hence, $(R_s,R_e)$ is an achievable
rate-equivocation pair through the BPSK-constrained Gaussian wiretap
channel.  Moreover, we note that the above lower bound is derived from
the Fano's inequality; thus it applies to any decoder at the
fictitious receiver. In fact, the value of the bound depends on the
choice of decoders only through $\epsilon_w$.  In the next section, we
perform computer simulation to estimate $\epsilon_w$ and then
employ~\eqref{eqn:nHMZn} to bound the equivocation rate achieved by
the proposed coding scheme as described above.  To get $\epsilon_w$, a
BP decoder is implemented for the fictitious receiver at the
wiretapper. In order to provide information about the secret message
$M$ to the BP decoder, the intrinsic log-likelihood ratios (LLRs) of
$c^k$ are explicitly set to $\pm \infty$ according to the true bit
values.


\section{Codes design and performance}\label{sec:irregldpc}

As mentioned in~\autoref{sec:intro}, reference~\cite{Klinc2009} uses a
systematic irregular LDPC code to encode the secret message $M$ (along
with some random bits) and then punctures the secret message bits in
the codeword prior to transmission in order to ``hide'' the secret
message from the wiretapper. The puncturing pattern is designed to
minimize the security gap.  Such a coding scheme can be viewed as an
unoptimized special case of our scheme proposed in
\autoref{sec:prop_cod}.  We show in this section that the
generalization in~\autoref{sec:prop_cod} allows us to systematically
optimize the irregular LDPC code for good secrecy performance.

To that end, let us apply the code search process proposed
in~\cite{CWWongTransInfoForensics10} to the present case.  Our
objective is to design the irregular LDPC code $\mathcal{C}'$ and a
puncturing scheme so that the secret LDPC code
$(\mathcal{C},\mathcal{W})$ works well for both the channel from the
source to the destination and the channel from the source to the
wiretapper (given the secret message). Let us first consider uniform
puncturing of the systematic bits of $\mathcal{C}'$, with $p$ denoting
the corresponding fraction of punctured variable nodes. Note that the
secret rate $R_s = \frac{p}{1-p}$. Also, write the rate of
$\mathcal{C}'$ as $R_c' = \frac{l}{m}$. Then $R_c'=\frac{R_c}{1+R_s}$.
For any fixed $R_s$, the discussion just below \eqref{eqn:nHMZn} at
the end of the previous section suggests that we should maximize
$R_c$, or equivalently $R_c'$, in order to maximize the achievable
equivocation rate.

Express the variable- and check-node degree distribution polynomials
of $\mathcal{C}'$ as, respectively, $\lambda(x) = \sum_{i=2}^{d_v}
\lambda_i x^{i-1}$ and $\rho(x) = \sum_{i=2}^{d_c} \rho_i x^{i-1}$,
where $\lambda_i (\rho_i)$ represents the fraction of edges emanating
from the variable (check) nodes of degree $i$ and $d_v$ ($d_c$) is the
maximum variable (check) degree.  For a fixed value of $R_s$ (which in
turn fixes $p$), the design objective is to then find $\lambda(x)$ and
$\rho(x)$ that maximize the code rate $R_c' = 1 - \frac{\int
  \rho(x) dx}{\int \lambda(x) dx}$ subject to the constraint that both
$\epsilon_d$ and $\epsilon_w$ vanish as the BP decoders iterate.

Fix $\rho(x)$, and let $e_d(\ell)$ and $e_w(\ell)$ denote the bit
error probabilities obtained by the BP decoders at the destination and
wiretapper, respectively, at the $\ell$th density evolution
iteration~\cite{RichardsonIT01_1,chung2001dld} when an initial
$\tilde\lambda(x) = \sum_{i=2}^{d_v} \tilde\lambda_i x^{i-1}$ is used.
Now, let $A_{\ell,j}$ denote the bit error probability obtained at the
destination by running the density evolution for $\ell$ iterations, in
which $\tilde\lambda(x)$ is used as the variable-node degree
distribution for the first $\ell - 1$ iterations and the variable-node
degree distribution with a singleton of unit mass at degree $j$ is
used for the final iteration. Let $B_{\ell,j}$ denote the similar
quantity for bit error probability obtained at the wiretapper. Then,
we have $e_d(\ell)=\sum_{j=2}^{d_v} A_{\ell,j} \tilde\lambda_j$ and
$e_w(\ell)=\sum_{j=2}^{d_v} B_{\ell,j} \tilde\lambda_j$.  Note that
the values of $A_{\ell,j}$ and $B_{\ell,j}$ are obtained via density
evolution. To account for the puncturing of $c^k$ at the destination's
BP decoder, the intrinsic LLR distribution entered into the density
evolution analysis for the destination's BP decoder is set to be a
mixture of the distribution of the channel outputs at the destination
and an impulse at $0$ with weights determined by the value of $p$.
Let $\epsilon>0$ be a small prescribed error tolerance. Suppose that
$\tilde \lambda(x)$ satisfies the property that $e_d(M_d) \leq
\epsilon$ and $e_w(M_w) \leq \epsilon$, for some integers $M_d$ and
$M_w$. Then, we can frame the $R_c$-maximizing code design problem as
the following linear program:
\begin{align*}
&  \max_{\lambda(x)} \sum_{j=2}^{d_v} \frac{\lambda_j}{j}
\mbox{~~subject to:}  \\
&  \sum_{j=2}^{d_v} \lambda_j = 1, ~~~
  \lambda_i \geq 0  &&\mbox{for~}2 \leq i \leq d_v \mbox{,} \\
&  \left |\sum_{j=2}^{d_v} A_{\ell,j} \lambda_j - e_s(\ell)\right|
  \leq \max[0, && \hspace*{-1.4em} \delta (e_s(\ell-1)-e_s(\ell))] \\ 
&  \hspace*{0.2in}\mbox{and~} \sum_{j=2}^{d_v} A_{\ell,j} \lambda_j \leq
e_s(\ell-1),  &&\mbox{for~}1 \leq \ell \leq M_s \\
&  \left |\sum_{j=2}^{d_v} B_{\ell,j} \lambda_j - e_w(\ell)\right|
\leq \max  [0, &&  \hspace*{-1.3em}\delta (e_w(\ell-1)-e_w(\ell))],  \\ 
&  \hspace*{0.2in}\mbox{and~} \sum_{j=2}^{d_v} B_{\ell,j} \lambda_j \leq
e_w(\ell-1),  &&\mbox{for~}1 \leq \ell \leq M_w, 
\end{align*}
where $d_v$ here is the maximum allowable degree of $\lambda(x)$ and
$\delta$ is a small positive number.  The solution $\lambda(x)$ of the
above linear program is then employed as the initial
$\tilde\lambda(x)$ for the next search round. The search process
continues this way until $e_d(M_d)$ or $e_w(M_w)$ becomes larger than
$\epsilon$, or until $\lambda(x)$ converges. We can also fix
$\lambda(x)$ and obtain a similar linear programming problem for
$\rho(x)$. The iterative search can then alternate between the linear
programs for $\lambda(x)$ and $\rho(x)$, respectively.

\begin{figure}
\centering
\begin{center}
\includegraphics[width=0.5\textwidth]{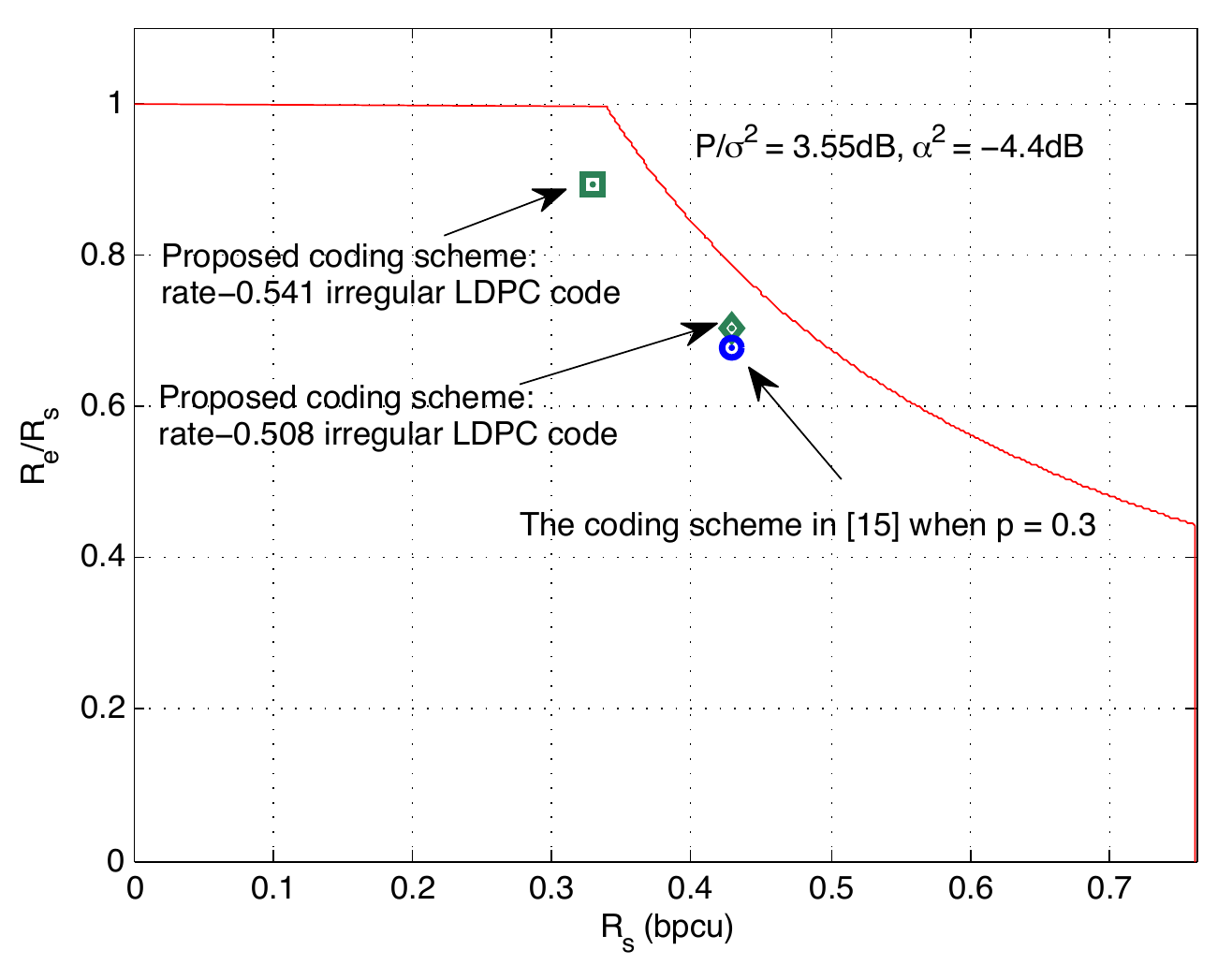}
\end{center}
\caption{Plot of $(R_s,\tilde{R}_e)$ pairs achieved by the proposed
  coding scheme and by the coding scheme in~\cite{Klinc2009} when
  $P/\sigma^2 = 3.55$~dB and $\alpha^2 = -4.4$~dB. The solid curve
  traces the boundary of the capacity-(fractional) equivocation region.}
\label{fig:irregular}
\vspace*{-10pt}
\end{figure}

For illustration, we apply the above code search procedure to two
different wiretap channel settings: (i) $P/\sigma^2 = 3.55$~dB and
$\alpha^2 = -4.4$~dB, and (ii) $P/\sigma^2 = 1.0$~dB and $\alpha^2 =
-1.0$~dB.  In both cases, the code search process starts with the
AWGN-optimized LDPC codes reported in~\cite{RichardsonIT01_2}.
Fig.~\ref{fig:irregular} shows the secrecy performance of a
rate-$0.541$ irregular LDPC code obtained by performing the code
search process with $R_s = 0.33$ under the first channel setting. The
degree distribution pair of this irregular LDPC code is shown
in~\autoref{tb:deg_dist_pair}.
\begin{table}
  \caption{Degree distribution pairs of the rate-$0.541$, rate-$0.508$, rate-$0.505$ irregular LDPC codes obtained from the proposed code search process.}
  \centering
  \begin{tabular}{|c||c|c|c|}
    \hline
    &rate-$0.541$ & rate-$0.508$ & rate-$0.505$\\
    \hline
    \hline
    $\lambda_2$ &  0.3013 & 0.2762 & 0.2599 \\
    \hline
    $\lambda_3$ &  0.1846 & 0.2804 & 0.2837 \\
    \hline
    $\lambda_4$ &  0.1510 & & 0.0281\\
    \hline
    $\lambda_9$ &  0.0614 & & \\
    \hline
    $\lambda_{10}$ & 0.3017 & 0.4434 & 0.4283\\
    \hline
    \hline
    $\rho_7$ & 0.3892 & 0.6086 & 0.6315\\
    \hline
    $\rho_8$ & 0.6054 & 0.3914 & 0.3532\\
    \hline
    $\rho_{10}$ & 0.0054 & & 0.0153\\
    \hline
  \end{tabular}
\label{tb:deg_dist_pair}
\vspace*{-20pt}
\end{table}
We obtain an instance of the irregular LDPC code by randomly
generating a bipartite graph which satisfies the two given degree
distributions.  The block length of the LDPC code is $n=10^6$, and all
length-$4$ loops are removed. Computer simulation is performed on this
code to estimate $\epsilon_d$ and $\epsilon_w$ as described
before. The estimated value of $\epsilon_w$ is employed to calculate
an achievable equivocation rate as in~\eqref{eqn:nHMZn}, provided that
$\epsilon_d \leq 0.01$ and $\epsilon_w \leq 0.01$. The resulting
achievable pair $(R_s,\tilde R_e)$ (where $\tilde R_e =
\frac{R_e}{R_s}$ is the fractional equivocation) is plotted against
the capacity-(fractional) equivocation region, whose boundary is shown
by the solid curve in the figure.  From Fig.~\ref{fig:irregular}, we
see that the pair $(R_s,\tilde{R}_e) = (0.33,0.89)$ (shown by the
square marker) is achieved by this rate-$0.541$ LDPC code.

\begin{figure}
\centering
\begin{center}
\includegraphics[width=0.5\textwidth]{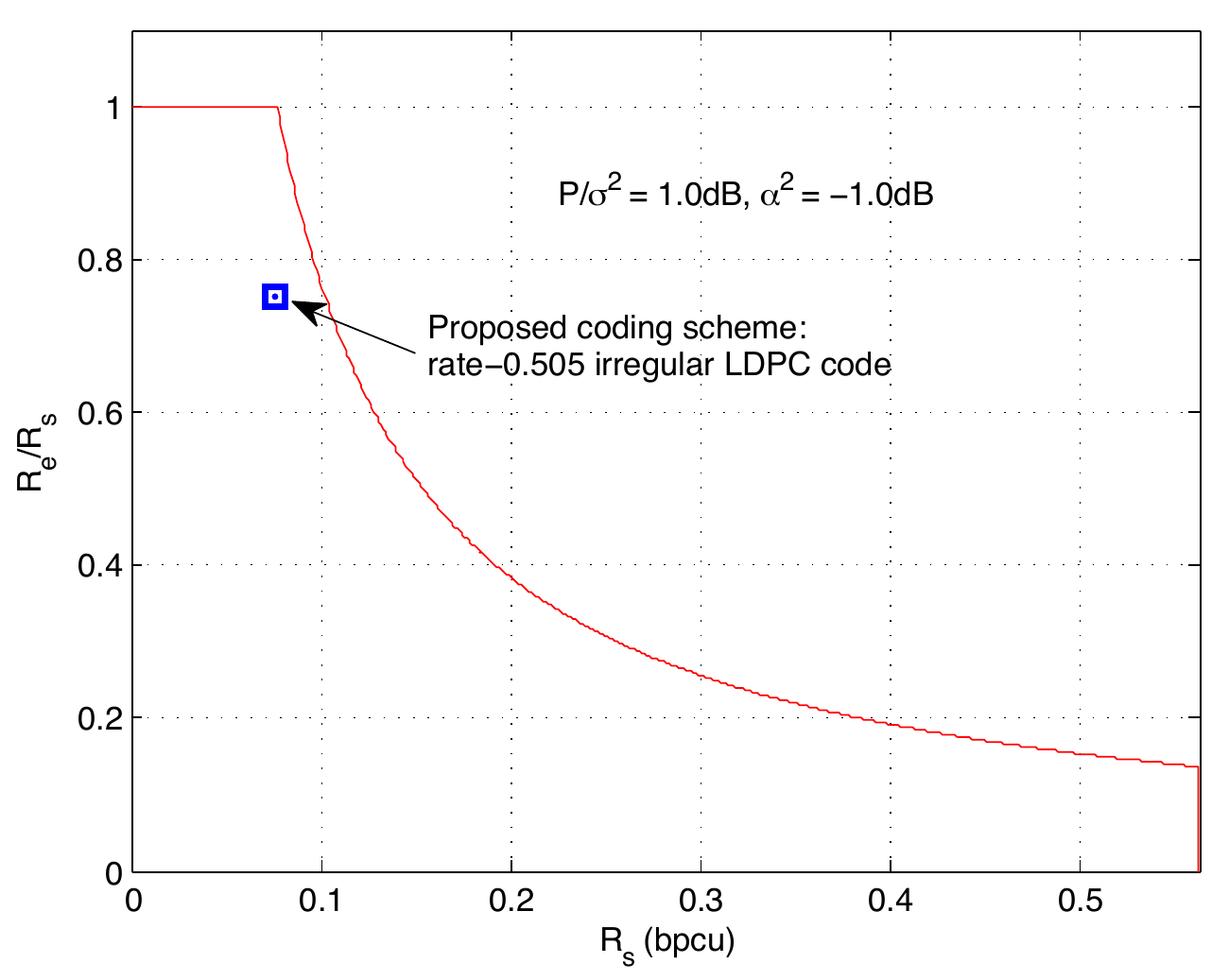}
\end{center}
\caption{Plot of the $(R_s,\tilde{R}_e)$ pair achieved by the proposed
  coding scheme when $P/\sigma^2 = 1.0$~dB and $\alpha^2 = -1.0$~dB.
  The solid curve traces the boundary of the capacity-(fractional) equivocation
  region.}
\label{fig:irregular_1dB}
\vspace*{-10pt}
\end{figure}

Next, we consider the more challenging case under the second channel
setting, in which the wiretapper's SNR is not much weaker than that of
the destination.  Fig.~\ref{fig:irregular_1dB} shows the secrecy
performance of a rate-$0.505$ irregular LDPC code obtained by
performing the code search process described above with $R_s = 0.076$.
The degree distribution pair of this irregular LDPC code can be found
in~\autoref{tb:deg_dist_pair}. We observe that the pair
$(R_s,\tilde{R}_e)=(0.076,0.76)$ (denoted by the square marker) is
achieved by this code.
In conclusion, the code search process described above gives irregular
LDPC codes with relatively good secrecy performance for different
values of $\alpha^2$. A similar code search process can also be
formulated to include optimization of the puncturing pattern. However,
we have not been able to obtain significantly better codes with the
modified search.

As mentioned before, the codes suggested in~\cite{Klinc2009} are
``unoptimized'' special cases of the coding scheme described here.  In
particular, a rate-$0.5$ irregular LDPC code with $p=0.3$ is employed
in~\cite{Klinc2009}, resulting in secret rate $R_s=0.43$.  The secrecy
performance of the coding scheme in~\cite{Klinc2009} is evaluated by
the security gap.
In our notation, that is to find the values $\tilde \beta$ and
$\alpha$ such that the decoding (bit) error probability of the secret
message at the destination is smaller than a prescribed value, and the
decoding (bit) error probability of the secret message at the
wiretapper is close to $0.5$.  The security gap is then defined as the
ratio of the SNR of the destination to that of the wiretapper,
i.e. $\frac{1}{\alpha^2}$.
As reported in~\cite{Klinc2009}, the security gap, with uniform
puncturing over all variable nodes of different degree for $p=0.3$ is
about $4.4$~dB.


To compare with our optimized codes, Fig.~\ref{fig:irregular} shows
the secrecy performance of the rate-$0.5$ code in~\cite{Klinc2009}
with $p=0.3$ evaluated by using~\eqref{eqn:nHMZn} as before under
channel setting (i). The pair $(R_s,\tilde{R}_e) = (0.43,0.68)$
(denoted by the circle marker) is achieved by this code.  We also
perform a code search under this channel setting with $R_s=0.43$ for
comparison. The pair $(R_s,\tilde{R}_e) = (0.43,0.70)$ (denoted by the
diamond marker) is achieved using the resulting rate-$0.508$ irregular
LDPC code.  We see that the irregular LDPC code obtained from the
proposed code search process also slightly outperforms the
``unoptimized'' one used in~\cite{Klinc2009} in terms of equivocation
rate.

Consulting back to Fig.~\ref{fig:cb}, we see that for
$\alpha^2=-4.4$~dB, the secrecy capacity of the BPSK-constrained
Gaussian wiretap channel never exceeds $0.34$~bpcu. Hence, the
fractional equivocation $\tilde R_e$ is straightly below $1$ at
$R_s=0.43$.  In fact, the highest achievable $\tilde{R}_e$ at
$R_s=0.43$ under this channel setting is only $0.78$
(cf. Fig~\ref{fig:irregular}). That means, we should not operate at
this rate if we target to achieve perfect secrecy.  In summary, the
proposed coding scheme and code search process provide a much more
systematic and flexible means to designing irregular LDPC codes for
the BPSK-constrained wiretap channel than the approach
in~\cite{Klinc2009}.

\section{Conclusions}\label{sec:con}
In this paper, we developed a coding scheme for sending secret
messages over the BPSK-constrained Gaussian wiretap channel. The
proposed coding scheme employs punctured systematic irregular LDPC
codes in which secret message bits are punctured.  To systematically
address the secret code design problem, we presented a
density-evolution based linear program to search for good irregular
LDPC codes to be used in the proposed coding scheme. Simulation
results showed that the irregular LDPC codes obtained from our search
can achieve secrecy performance relatively close to the boundary of
the capacity-equivocation region of the BPSK-constrained Gaussian
wiretap channel.

\end{document}